\documentclass[aps,prl,reprint,superscriptaddress]{revtex4-2}

\usepackage[colorlinks=true, allcolors=blue]{hyperref}
\usepackage{hypernat}
\usepackage{orcidlink}
\usepackage{bm}
\usepackage{amsfonts}
\usepackage{amsmath}
\usepackage{amssymb}
\usepackage{graphicx}	
\usepackage{booktabs}
\usepackage{diagbox}
\usepackage{makecell}
\usepackage{mathptmx}

\newcommand{\rtm}{r_{\mathrm{200m}}}

\newcommand{\Mtm}{M_{\mathrm{200m}}}

\newcommand{\Msol}{\mathrm{M}_{\odot}}

\newcommand{\entropunit}{{\rm keV\,cm^{2}}}

\begin{document}

\title{\textbf{Thermostats, Not Engines: A New Picture of Halo Gas Regulation} }%

\author{Hiranya~V.~Peiris$^{\orcidlink{0000-0002-2519-584X}}$}
\email[]{hiranya.peiris@ast.cam.ac.uk}
\affiliation{Institute of Astronomy and Kavli Institute for Cosmology, University of Cambridge, Madingley Road, Cambridge, CB3 0HA, UK}
\affiliation{Cavendish Laboratory, Department of Physics, University of Cambridge, JJ Thomson Avenue, Cambridge, CB3 0HE, UK}
\affiliation{The Oskar Klein Centre, Department of Physics, Stockholm University, AlbaNova University Centre, 106 91 Stockholm, Sweden}

\author{Andrew~Pontzen$^{\orcidlink{0000-0001-9546-3849}}$}
\email[]{andrew.p.pontzen@durham.ac.uk}
\affiliation{Institute for Computational Cosmology, Department of Physics, Durham University, South Road, Durham, DH1 3LE, UK}

\author{Madalina~N.~Tudorache$^{\orcidlink{0000-0002-7288-6627}}$}
\affiliation{Institute of Astronomy and Kavli Institute for Cosmology, University of Cambridge, Madingley Road, Cambridge, CB3 0HA, UK}

\author{Anik~Halder$^{\orcidlink{0000-0002-0352-9351}}$}
\affiliation{Institute of Astronomy and Kavli Institute for Cosmology, University of Cambridge, Madingley Road, Cambridge, CB3 0HA, UK}

\author{Stephen~Thorp$^{\orcidlink{0009-0005-6323-0457}}$}
\affiliation{Institute of Astronomy and Kavli Institute for Cosmology, University of Cambridge, Madingley Road, Cambridge, CB3 0HA, UK}

\author{Sinan~Deger$^{\orcidlink{0000-0003-1943-723X}}$}
\affiliation{Institute of Astronomy and Kavli Institute for Cosmology, University of Cambridge, Madingley Road, Cambridge, CB3 0HA, UK}

\author{Joop~Schaye$^{\orcidlink{0000-0002-0668-5560}}$}
\affiliation{Leiden Observatory, Leiden University, PO Box 9513, 2300 RA Leiden, the Netherlands}

\author{Matthieu~Schaller$^{\orcidlink{0000-0002-2395-4902}}$}
\affiliation{Leiden Observatory, Leiden University, PO Box 9513, 2300 RA Leiden, the Netherlands}
\affiliation{Lorentz Institute for Theoretical Physics, Leiden University, PO Box 9506, 2300 RA Leiden, the Netherlands}

\date{\today}

\begin{abstract}
We propose that black hole feedback regulates gas in massive halos by establishing an entropy ceiling; the resulting buoyant gas migrates to the virial radius with no additional energy input required. The FLAMINGO simulations support this picture: at the virial radius, outflow entropy is mass-independent for isotropic thermal feedback but depends on the solid angle of directly heated gas for jet feedback. Above a critical halo mass $M_\mathrm{crit} \approx 10^{13.5\text{--}14}\, \Msol$, virial shocks overwhelm the ceiling, predicting rejuvenation of star formation in the most massive galaxies, supported by new low-redshift evidence from star formation rates and morphologies.
\end{abstract}

\maketitle


\textit{Introduction.}---The regulation of gas cooling in massive halos is commonly attributed to continuous energy injection from active galactic nuclei \cite[AGN; for a review, see][]{Fabian2012}. In this picture, the black hole acts as an engine, actively powering outflows that offset radiative cooling throughout the halo. We propose an alternative picture in which AGN feedback establishes an entropy ceiling---a mass-independent upper limit on the entropy that it can imprint---whose value is set by local black hole physics. Gravitational dynamics---sloshing driven by mergers and accretion---then transports this buoyant gas outward. The kinetic and thermal energy of the outflow is set in the inner halo and conserved during transport to the virial radius. The black hole acts as a thermostat rather than an engine: it sets the entropy in the inner halo, and buoyancy transports the heated gas outward.

We develop this framework using the FLAMINGO
cosmological hydrodynamical simulations~\cite{schaye23, kugel23}. The companion paper by Pontzen et al.\ (hereafter PP26) presents measurements of gas inflows and outflows as a function of radius and halo mass, demonstrating that AGN energy injection at small radii accounts for the difference between inflow and outflow energy rates. It shows that an entropy ceiling, whose properties depend on the feedback implementation, is established via resolved shock heating in the inner halo. While the thermostat picture is developed here using FLAMINGO, the framework should hold generically in any simulation where feedback energy is sufficient to establish a buoyant entropy ceiling. We test it against FLAMINGO variants with differing AGN feedback strength and geometry (thermal vs.\ jet), and confront its predictions with new observational evidence for rejuvenation of star formation in massive galaxies at low redshift.

\textit{The entropy thermostat.}---Our work is motivated by Lucie-Smith et al.~\cite{lucie-smith2025}, who showed that baryonic feedback in FLAMINGO most efficiently redistributes gas when halos reach an instantaneous mass $M_{200\mathrm{m}} \approx 10^{12.8}\,\Msol$, with this characteristic scale approximately independent of redshift. We show in PP26 that in the FLAMINGO simulations, the kinetic and thermal energy rates at $\rtm$ scale linearly with halo mass: $\dot{E}_{\rm outflow} \propto M$. This result holds across group-scale halos ($M \approx 10^{12.5}$--$10^{13.5}\,\Msol$), above the threshold for stable virial shocks~\cite{DekelBirnboim2006} and within the mass range where AGN feedback is expected to be most influential. This finding is approximately independent of feedback implementation (thermal vs.\ jet). Cosmological mass accretion delivers gas at a rate $\dot{M}_{\rm inflow} \propto M$, and the absence of sharp transitions between inflow- and outflow-dominated regimes in the FLAMINGO gas fractions requires approximate equilibrium, and therefore $\dot{M}_{\rm outflow} \propto M$. The outflow energy rate is $\dot{E}_{\rm outflow} = \varepsilon_{\rm out}\,\dot{M}_{\rm outflow}$, where $\varepsilon_{\rm out}$ is the specific energy of outflowing gas.  For $\dot{E}_{\rm outflow} \propto M$ to hold, $\varepsilon_{\rm out}$ must be mass-independent. X-ray observations of galaxy groups have long shown entropy in excess of self-similar expectations~\cite{Ponman1999}, later refined to show that the entropy increases with radius out to at least several hundred kpc~\cite{Panagoulia2014}. The mass-independent ceiling we identify here is a distinct quantity, measured at the virial radius as a function of halo mass. Here we ask, what establishes this ceiling?

\begin{figure*}
\includegraphics[width=\textwidth]{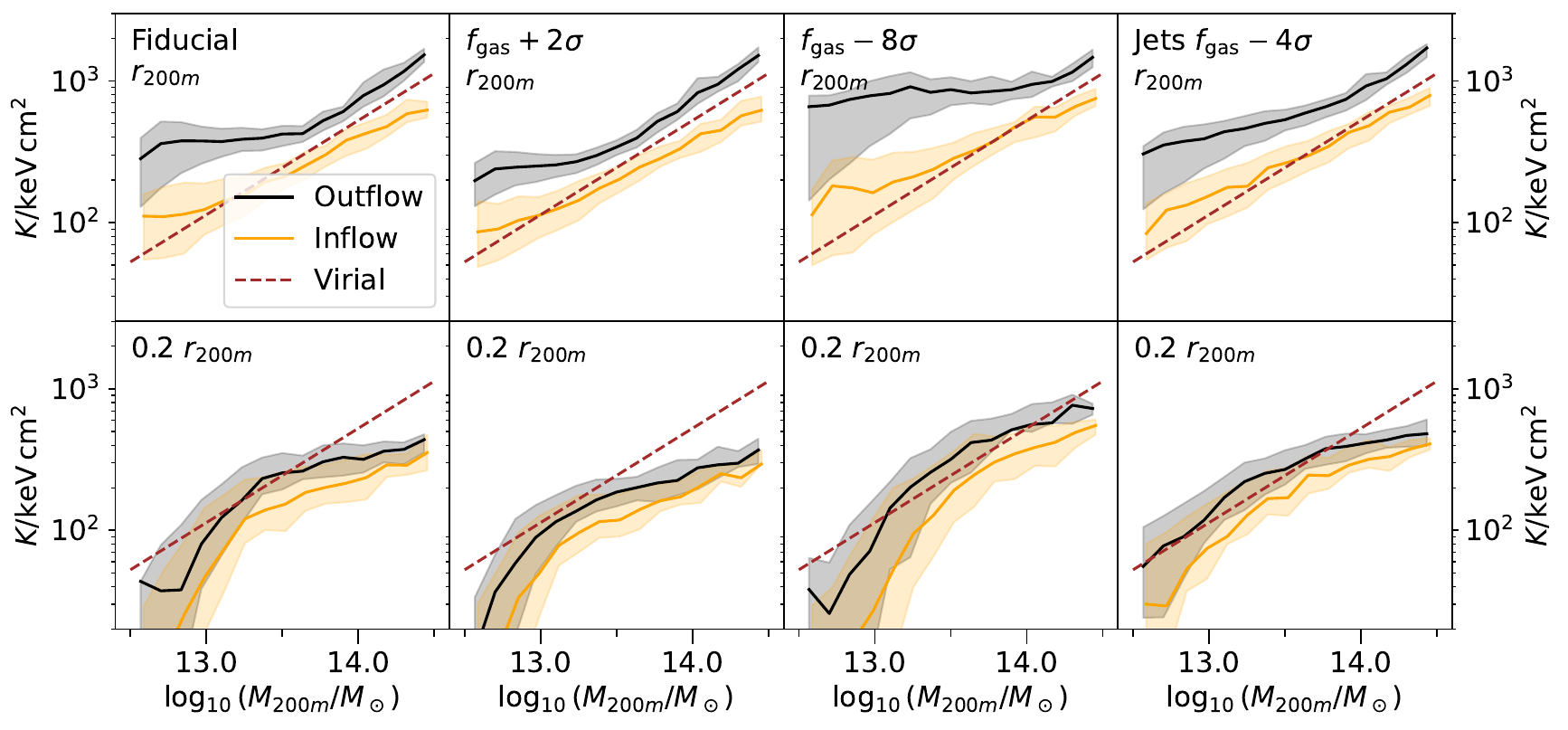}
\caption{Entropy as a function of halo mass at $\rtm$ (top) and $0.2\,\rtm$ (bottom) for four FLAMINGO AGN variants at $z=0.4$, from left to right: fiducial, weak AGN (fgas$+2\sigma$), strong AGN (fgas$-8\sigma$), and jets (fgas$-4\sigma$). Lines show flow-weighted mean entropy for outflowing (black) and inflowing (orange) gas; shaded bands enclose the 16--84th percentile range across the halo population. The red dashed line shows virial scaling $K \propto M^{2/3}$, where the normalization is computed from first principles with no free parameters (see PP26).}
\label{fig:entropy}
\end{figure*}

Three possibilities present themselves. First, the ceiling could arise from purely gravitational physics---perhaps some self-regulating thermalization process that happens to produce constant entropy at $\rtm$. This seems unlikely: gravitational shock heating naturally produces $K \propto M^{2/3}$, and there is no obvious mechanism to flatten this scaling. Second, the ceiling could be set by pre-heating in the cosmic web before gas enters the halo, imprinting a mass-independent entropy from the intergalactic medium. However, Lagrangian particle tracking in the OWLS
simulations \cite{Schaye2010, McCarthy2011} showed that the elevated gas entropy in present-day groups was imprinted by AGN feedback in low-mass progenitor halos at $z \approx 2$--$4$~\cite{McCarthy2011}---effectively ruling out this possibility. The remaining option is that AGN feedback establishes the ceiling by giving rise to high-entropy gas that rises buoyantly to $\rtm$.

We argue for the third possibility, but not in the usual sense of AGN directly powering outflows at large radii. Instead, the AGN injects energy near the black hole at a rate governed by subgrid parameters that are independent of halo mass. In FLAMINGO's thermal prescription~\cite{booth09}, accreted rest-mass energy heats $n_{\mathrm{heat}}$ neighboring particles by a fixed $\Delta T_{\mathrm{AGN}}$; in the jet variant~\cite{husko22,kugel23}, it is injected kinetically within $7.5^\circ$ half-opening-angle cones at a fixed velocity $v_{\mathrm{jet}}$. The entropy generated per feedback event is therefore approximately mass-independent (PP26 derives the resulting entropy ceiling from the Rankine--Hugoniot shock relations). The entropy injected depends on the subgrid feedback parameters, but these are calibrated to observed gas fractions as a function of halo mass~\cite{kugel23}, anchoring the ceiling to observational constraints. This high-entropy gas is buoyant: it rises outward through the halo atmosphere until it reaches a radius where its entropy matches that of the surrounding gas. In halos below a critical mass $M_{\rm crit}$, this settling radius lies in the outer halo, where AGN-heated gas accumulates and establishes the observed entropy ceiling. Cosmological infall drives sloshing dynamics---the partial thermalization and turnaround of infalling gas---on top of this entropy boundary condition.

This mechanism implies that the AGN signature appears at large radii rather than small radii. At inner radii, any gas heated by the AGN promptly floats away, leaving behind fresh material continuously replenished by infall and cooling. The inner halo therefore should display the self-similar entropy profile $K \propto M^{2/3}$, characteristic of gravitational physics. At $\rtm$, by contrast, the buoyantly transported gas should pile up, creating a mass-independent entropy boundary condition.

Buoyancy has long been recognised as central to gas transport in groups and clusters~\cite{Voit2005, Fabian2012}. Bower et al.~\cite{Bower2017} applied a buoyancy-failure criterion to supernovae-driven outflows to explain the onset of AGN feedback at $\sim 10^{12}\,\Msol$; our picture is complementary, applying at higher masses ($\Mtm \gtrsim 10^{12.5}\,\Msol$) where AGN feedback itself ceases to be effective because the virial entropy overwhelms the ceiling.

\textit{Tests using the FLAMINGO simulations.}---Figure~\ref{fig:entropy} (left column; see also PP26) shows entropy as a function of halo mass at two radii for the fiducial FLAMINGO simulation at $z=0.4$. At $\rtm$, the entropy of outflowing gas is approximately constant at $K = 370\, \entropunit$ across the range $10^{12.5} < \Mtm / \Msol < 10^{13.5} $.  The entropy of inflowing gas rises with mass, approximately following the virial scaling $K \propto M^{2/3}$. This contrast between inflow and outflow properties at the same radius points to distinct thermal histories: inflowing gas is fresh material from the cosmic web, while outflowing gas has been processed by AGN feedback.

At $\rtm$, the entropy ceiling is fully established. The transition from AGN-dominated to gravitationally-dominated thermodynamics occurs gradually with decreasing radius. At intermediate radii ($0.6\,\rtm$), the scaling is intermediate, with the entropy ceiling weakening as gravitational physics emerges. By $0.2\,\rtm$ (Fig.~\ref{fig:entropy}, bottom left), the transition is complete: both inflow and outflow energy rates follow the self-similar scaling $\dot{E} \propto M^{5/3}$, and entropy rises with mass as $\propto M^{2/3}$. At this inner radius, $\dot{E}_{\rm inflow} \approx \dot{E}_{\rm outflow}$, indicating approximate energy balance: the inner halo acts as a ``reflector'', returning buoyant gas outward without net energy deposition. The halo thus exhibits a continuous transition from AGN-controlled thermodynamics in the outer regions to purely virial dynamics in the interior.

The framework predicts a critical mass above which the virial entropy exceeds the AGN-set ceiling: self-similar scaling is restored even at $\rtm$, and $\dot{E}_{\rm outflow}$ steepens from $\propto M$ to $\propto M^{5/3}$. In the fiducial simulation, the outflow and inflow entropies converge at $M_{\rm crit} \approx 10^{13.5\text{--}14}\,\Msol$ (Fig.~\ref{fig:entropy}), coinciding with the mass scale at which FLAMINGO shows gas fractions recovering toward the cosmic mean.

\textit{Varying AGN feedback strength.}---If the entropy ceiling is established by buoyant transport of AGN-heated gas, then varying the AGN heating temperature $\Delta T_{\rm AGN}$ in the fiducial FLAMINGO thermal AGN feedback simulation should produce predictable changes. Stronger feedback should raise the entropy ceiling, extend its radial reach inward, and push $M_{\rm crit}$ to higher masses. Weaker feedback should lower the ceiling, shrink its radial extent, and reduce $M_{\rm crit}$---potentially to the point where the ceiling vanishes entirely within the observed mass range.

FLAMINGO provides two relevant variants for their fiducial thermal AGN feedback implementation: a weak AGN model (fgas$+2\sigma$), and a strong AGN model (fgas$-8\sigma$). The results confirm the framework's predictions in striking detail (Fig.~\ref{fig:entropy}, second and third columns). In the strong AGN simulation, the entropy ceiling at $\rtm$ is elevated to $K \approx 800\, \entropunit$, roughly double the fiducial value, and $M_{\rm crit}$ extends to $\Mtm \approx 10^{14}\, \Msol$. The transition zone shifts inward: at $0.6\,\rtm$, the strong AGN run displays a nearly flat entropy ceiling resembling the fiducial behavior at $\rtm$, though gravitational scaling persists at $0.2\,\rtm$ even with enhanced feedback. The weak AGN simulation shows the opposite: the ceiling has vanished entirely, with virial scaling $K \propto M^{2/3}$ and $\dot{E} \propto M^{5/3}$ at all radii and $M_{\rm crit} \lesssim 10^{12.5}\,\Msol$---the entire halo behaves as if AGN feedback were absent. If gravitational physics alone could produce a mass-independent entropy ceiling, it would persist regardless of AGN strength---yet the weak AGN case shows virial scaling fully restored. The ceiling requires AGN feedback above a threshold strength, and the AGN heating temperature directly controls both its amplitude and radial extent.

\textit{Thermal vs.\ jet feedback: geometry matters.}---The preceding analysis uses FLAMINGO's fiducial thermal feedback, in which AGN energy is deposited isotropically. FLAMINGO also provides jet feedback variants~\cite{husko22, kugel23} in which energy is instead deposited kinetically along the black hole spin axis. We analyze the jet fgas$-4\sigma$ variant, the most appropriate comparison to the fiducial thermal case (see PP26 for details). Comparing these implementations tests whether the entropy ceiling is specific to isotropic injection or a more general consequence of feedback-driven buoyant transport.

Three features of the jet entropy scaling require explanation (Fig.~\ref{fig:entropy}). First, at $\rtm$ the jet case shows $K \propto M^{0.22}$ for $\Mtm \lesssim 10^{13.5}\,\Msol$, rather than the flat ceiling seen in the thermal cases. Second, this mass scaling is the same at $\rtm$ and $2\,\rtm$ (see Appendix and Fig.~\ref{fig:entropy_2r200m}). Third, the transition to virial scaling occurs at the same critical mass $M_{\rm crit} \approx 10^{13.5}\,\Msol$ as in the thermal case. A further diagnostic sharpens the interpretation: we verify that the outflow temperature scaling with halo mass is essentially identical between the thermal and jet cases at both radii, despite the very different entropy behavior. Since $K = T/\rho^{2/3}$, the entropy difference between the two implementations is entirely a density effect at the particle level, not a temperature effect. This points to the geometry of the outflow, rather than the energetics of the heating mechanism.

All three observations follow from a single physical picture. In the thermal case, the isotropic shock processes gas at all solid angles, so every line of sight through a stacked sample sees directly processed, high-entropy gas; the result is a mass-independent ceiling. In the jet case, each halo has a bipolar outflow that imprints entropy on the surrounding halo gas covering a limited solid angle. Although the jet-inflated bubble broadens substantially as it rises buoyantly---simulation visualizations confirm that jet and thermal bubbles are morphologically similar at large scales \cite{ondaro-mallea25}---it does not become fully isotropic. When we stack over many halos with random jet orientations, the median outflow entropy at $\rtm$ is therefore a mix of two components: gas within the broadened bubble that has been directly processed, and gas outside it that sits near the virial entropy.

The mass scaling is preserved across radii because the entropy of both components is set in the halo interior and conserved during buoyant transport outward, and the angular structure of the bubble is similarly preserved. The transition to virial scaling occurs at the same mass because $M_{\rm crit}$ is determined by when the virial entropy exceeds the feedback-imprinted entropy, and at that mass the thermal and jet entropies have converged.

A simple two-component model makes this quantitative. Writing the stacked entropy as
\begin{equation}
    K_{\rm stack}(M) = f\, K_d + (1-f)\, K_{\rm vir}(M),
    \label{eq:mixing}
\end{equation}
where $f$ is the directly processed fraction at mass-independent entropy $K_d$ and $(1-f)$ sits at virial entropy, we can solve for $f$ from the data. Over the mass range $10^{12.5}$--$10^{13.5}\,\Msol$, the virial entropy increases by a factor $10^{2/3} \approx 4.6$ while the measured jet entropy increases by $10^{0.22} \approx 1.7$. From the FLAMINGO data we take the thermal ceiling value $K_d \approx 370\,\entropunit$ and $K_{\rm vir}(10^{12.5}\,\Msol) \approx 53 \, \entropunit$. Requiring the ratio $K_{\rm stack}(10^{13.5}\,\Msol) / K_{\rm stack}(10^{12.5}\,\Msol) = 1.7$ then yields $f \approx 0.4$. The corresponding solid angle for two opposing cones, $f_\Omega = 1 - \cos\theta$, implies an effective half-angle of $\sim 50^\circ$---a factor of seven broader than the $7.5^\circ$ injection angle. PP26 confirms that shock energy injection is more collimated in the jet case than in the thermal case.

The jet case shows that the thermostat picture generalizes beyond isotropic thermal injection. What changes is the geometry: isotropic processing gives a flat ceiling, while anisotropic processing gives a mass scaling whose value encodes the solid angle of the directly heated region.

\begin{table*}
    \centering
    \caption{Approximate scaling relations across halo radii in the FLAMINGO simulations at $z=0.4$. The entropy scaling at $r_\mathrm{200m}$ below $M_\mathrm{crit}$ depends on feedback geometry: isotropic thermal injection gives $K \approx \mathrm{const}$, while anisotropic jet injection gives $K \propto M^{0.22}$. Energy rates $\dot{E}$ refer to kinetic and thermal energy flux, excluding gravitational potential energy. This choice reflects the physical picture: outflowing gas is buoyant, and its kinetic and thermal energy is set in the inner halo and conserved during transport to $r_\mathrm{200m}$ (demonstrated in PP26).} 
    \label{tab:scalings}
    \begin{tabular}{lcccc}
        \toprule\toprule
         \backslashbox{\textbf{Quantity}}{\textbf{Radius}} & \makecell{\textbf{Inner halo} \\ \textbf{($0.2\,\rtm$)}} & \makecell{\textbf{Transition zone} \\ \textbf{($0.6\,\rtm$)}} & \makecell{\textbf{Outer halo} \\ \textbf{($\rtm$, $M < M_{\rm crit}$)}} & \makecell{\textbf{Outer halo} \\ \textbf{($\rtm$, $M > M_{\rm crit}$)}}\\
        \midrule
        Dominant physics & Virial & Mixed & AGN ceiling & Virial \\
        $K_{\rm outflow}$ & $\propto M^{2/3}$ & Intermediate & $\text{const}$ (thermal) / $\propto M^{0.22}$ (jets) & $\propto M^{2/3}$ \\
        $\dot{E}_{\rm outflow}$ & $\propto M^{5/3}$ & Intermediate & $\propto M$ & $\propto M^{5/3}$ \\
        \bottomrule\bottomrule
    \end{tabular}
\end{table*}

Finally, in the jet case, the gas fraction interior to $\rtm$ is suppressed to roughly a quarter of the cosmic baryon fraction at $M_{200\mathrm{m}} \simeq 10^{12.5\text{--}13.0}\,\Msol$, recovering toward the cosmic mean by $\sim M_\mathrm{crit}$~\cite{lucie-smith2025}. McDonald et al.~\cite{McDonald2018} found that in cool-core systems---which typically host jet-mode AGN---star formation rates are uncorrelated with cooling rates in groups but correlated in clusters: below $M_\mathrm{crit}$, the depleted interior decouples star formation from cooling; above it, gas reaches the core where shorter cooling times permit condensation.

\textit{Discussion.}---The picture that emerges cleanly separates the roles of AGN feedback and gravity, and generalizes across feedback implementations (the jet case shows larger scatter, presumably due to orientation dependence of individual outflows). Table~\ref{tab:scalings} summarizes the scaling relations across radii.

This unified physical picture is further supported by Hu\v{s}ko et al.~\cite{husko24}, who found that thermal, kinetic jet, and hybrid feedback implementations in idealized group and cluster simulations all produce entropy plateaus at group scales. Braspenning et al.~\cite{braspenning24} showed that FLAMINGO thermal and jet variants, calibrated to the same stellar mass function and cluster gas fraction data, yield similar gas fractions at large radii but remain distinguishable in their core properties---consistent with the ceiling controlling large-scale gas content regardless of how the energy was deposited.

Our measurements provide a complementary, Eulerian perspective on McCarthy et al.~\cite{McCarthy2011}, who used Lagrangian tracking to show that elevated group entropy was imprinted by AGN in low-mass progenitors at $z \approx 2$--4. Whether the ceiling was established by early ejection or is maintained by ongoing buoyant transport---or both---the observable consequence is the same, and the weak-AGN variant demonstrates that it requires feedback above a threshold strength. Altamura et al.\ \cite{Altamura2025} traced the formation of entropy plateaus in individual halos reaching $\sim 10^{12}\,\Msol$; our population-level measurements establish how the ceiling depends on halo mass, radius, feedback strength, and geometry at fixed epoch.

\textit{Testable predictions.}--- What is the fate of the gas that remains buoyant above the entropy ceiling? PP26 shows that in the intermediate mass regime ($10^{13}$--$10^{13.7}\,\mathrm{M}_\odot$), outflowing gas lingers just outside $\rtm$, with densities that do not decline despite net outflows. This reservoir re-enters the virial radius as halos grow past $M_{\rm crit}$, raising the potential for rejuvenation of star formation in the most massive galaxies.

X-ray observations of galaxy clusters establish that cooling times in cluster cores are short: within the inner $\sim 50\,{\rm kpc}$, strong cool-core clusters have $t_{\rm cool} \lesssim 1\,{\rm Gyr}$, well below the Hubble time~\cite{Cavagnolo2009, Hudson2010}. At group scales ($M \approx 10^{13}\,\Msol$), the outflow energy rate exceeds the inflow rate at $0.2\,\rtm$ (PP26), indicating heating-dominated conditions where quenching is maintained. At cluster scales ($M \gtrsim 10^{14}\,\Msol$), outflow and inflow rates become comparable; cooling competes with heating, and some gas can condense to fuel star formation. The transition occurs at $M_{\rm crit} \approx 10^{13.5\text{--}14}\,\Msol$.

The FLAMINGO stellar-to-halo mass relation~\cite{schaye23} maps $M_\mathrm{crit} \approx 10^{13.5\text{--}14}\,\Msol$ to a stellar mass $M_* \approx 10^{11.2\text{--}11.5}\,\Msol$. Deger et al.\ \cite{deger25} recently presented quenched fractions in massive galaxies using a population-level inference approach \cite[\textsc{pop-cosmos};][]{alsing24, thorp25b} which self-consistently models and accounts for selection effects. Intriguingly, they found that the quenched fraction of massive galaxies declines at $M_* \gtrsim 10^{11}\,\Msol$. This finding coincides with the mass range where our framework predicts the possibility of rejuvenation. 

\begin{figure}
\includegraphics[width=\columnwidth]{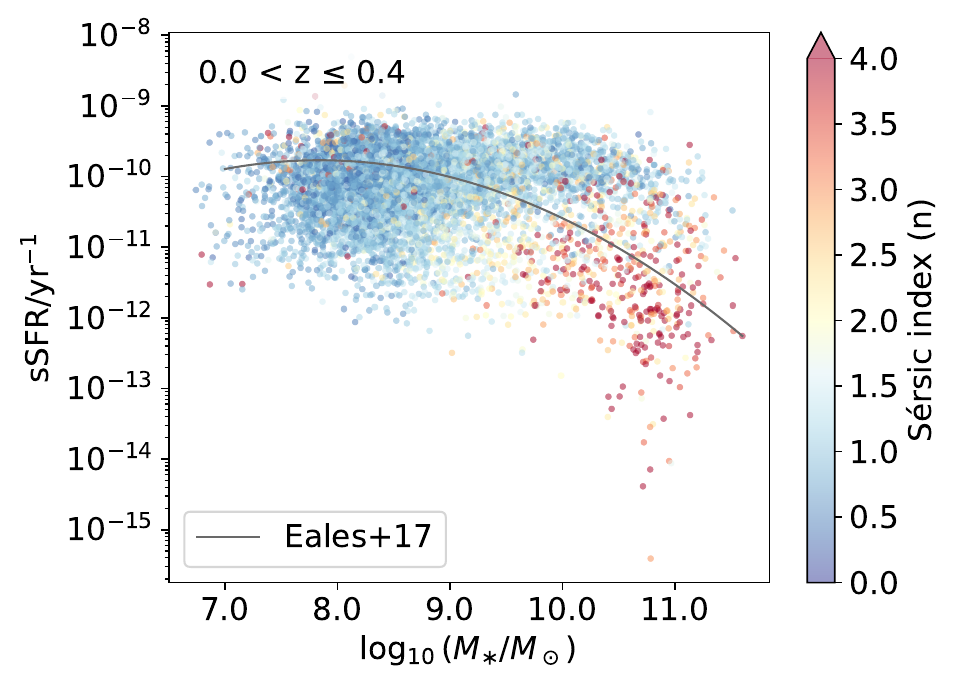}
\caption{Specific star formation rate as a function of stellar mass, inferred using \textsc{pop-cosmos}, for the COSMOS2020 galaxies crossmatched to COSMOS-Web in the $0.0 < z < 0.4$ redshift interval, color-coded by S\'{e}rsic index. The dark gray line shows the relation in Eales et al.~\cite{eales17}.}
\label{fig:m-ssfr-morpho}
\end{figure}

Our results imply that the most massive galaxies show declining quenched fractions not because feedback has failed, but because the energy balance has shifted from heating-dominated to marginal. Eales et al.\ \cite{eales17, eales18b} showed that galaxies follow a ``galaxy end sequence''---a continuous, curved sequence in specific star formation rate vs.\ stellar mass with no clear break between star-forming and quiescent populations at high masses---results that require more gradual evolutionary processes than rapid quenching \cite{quilley22}. In Fig.~\ref{fig:m-ssfr-morpho}, we present an updated version of this analysis using $9319$ galaxies in the $0.0<z<0.4$ redshift interval. Their stellar masses and specific star formation rates are inferred from $26$-band COSMOS2020 photometry~\cite{weaver22} using \textsc{pop-cosmos}~\cite{thorp24, thorp25b}, and S\'{e}rsic indices are crossmatched from COSMOS-Web~\cite{casey23, shuntov25}. This represents a substantial advance over previous work: conventional template fitting is replaced with Bayesian parameter inference using a calibrated prior \cite[\textsc{pop-cosmos};][]{alsing24, thorp25b} and 16-parameter emulated stellar population synthesis model \cite[\textsc{Speculator};][]{alsing20}, constrained by broad ($\sim0.35$--$4.4$~\textmu{}m) wavelength coverage; and morphological measurements are derived from joint modeling of four-band restframe near-infrared JWST NIRCam imaging, providing a more robust structural proxy than either visual classification or ground-based decomposition.

In this sample, we see a gradual transition between the quiescent and the star-forming galaxies, consistent with the galaxy end sequence~\cite{eales17, eales18b, quilley22}. Furthermore, the S\'{e}rsic index $n$ also evolves gradually along this curve, with even the highest mass galaxies showing a broad range of morphologies. In particular, we see that disk-like ($n \leq 2$) morphologies persist in the most massive galaxies, where quenched, bulge-dominated systems are expected to predominate. This is behavior expected above $M_\mathrm{crit}$; rather than a sharp division between star-forming and quenched populations, massive galaxies occupy the full range of star formation rates and structural types. 

\textit{Conclusion.}---We have shown that AGN feedback regulates halo gas by establishing an entropy boundary condition rather than by directly powering outflows. The FLAMINGO simulations with varying feedback strength and geometry confirm that the picture holds generically: the entropy ceiling emerges whenever feedback is sufficiently strong, its mass scaling encodes the solid angle of the directly heated region, and a critical mass marks where virial shocks overwhelm the ceiling. Observational signatures at high stellar masses at low redshift (declining quenched fractions and a broad range of morphologies) are consistent with the predicted rejuvenation above the critical mass scale. Once the entropy ceiling is set, it persists until halos grow past the critical mass---implying that quenching is not permanent, and that the most massive galaxies may be the first to escape it.

\begin{acknowledgments}
\textit{Acknowledgements.}---We thank Alex Amon, Leah Bigwood, George Efstathiou, Andy Fabian, Simone Ferraro, Boryana Hadzhiyska, Luisa Lucie-Smith, Ian McCarthy, Daisuke Nagai, and Emmanuel Schaan for helpful conversations.  This work has been supported by funding from the European Research Council (ERC) under the European Union's Horizon 2020 research and innovation programmes (grant agreement no.\ 101018897 CosmicExplorer and no.\ 818085 GMGalaxies), and the research project grant ``Understanding the Dynamic Universe'' funded by the Knut and Alice Wallenberg Foundation under Dnr KAW 2018.0067. HVP was additionally supported by the G\"{o}ran Gustafsson Foundation for Research in Natural Sciences and Medicine. HVP and AP thank the organizers of the inspiring {\it Cosmic Ecosystems} workshop held in July 2025 at the Perimeter Institute, where this work was initiated, for their kind hospitality.  This research was supported in part by Perimeter Institute for Theoretical Physics. Research at Perimeter Institute is supported by the Government of Canada through the Department of Innovation, Science, and Economic Development, and by the Province of Ontario through the Ministry of Colleges and Universities.

\textit{Author contributions.}---{\bf HVP}: conceptualization; formal analysis; investigation; methodology; validation; visualization; writing -- original draft, review \& editing. {\bf AP}: conceptualization; formal analysis; methodology; software; validation; visualization; writing -- review \& editing. {\bf MNT}: investigation; writing -- review \& editing. {\bf AH}: investigation; writing -- review. {\bf ST}: writing -- review \& editing. {\bf SD}: writing -- review \& editing. {\bf JS}: project administration (FLAMINGO); data curation; resources; writing -- review \& editing. {\bf MS}: data curation; resources.

\textit{Declaration of LLM use.}---Anthropic's Claude Opus 4.6 has been used to assess flow and redundancy of text, obtain suggestions for condensing, and proofreading.

\end{acknowledgments}

\bibliography{apssamp} 

\bibsection 

\appendix

\begin{figure*}
\includegraphics[width=\textwidth]{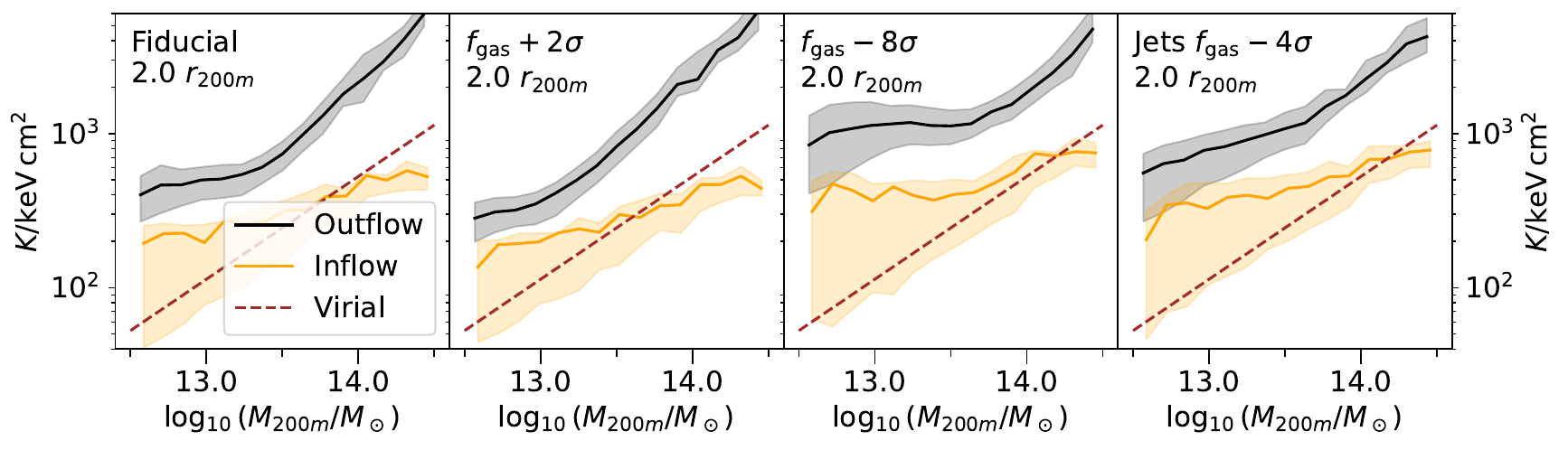}
\caption{Entropy as a function of halo mass at $2\,r_\mathrm{200m}$ for the same four FLAMINGO AGN variants and with the same line styles as in Fig.~\ref{fig:entropy}. The jet entropy slope $K \propto M^{0.22}$ is preserved at $2\,r_\mathrm{200m}$ (compare the right panel of Fig.~\ref{fig:entropy}), while the fiducial thermal case develops a positive slope and falls below the jet case at low masses---a reversal of the ordering at $r_\mathrm{200m}$. The strong AGN case (fgas$-8\sigma$) maintains a nearly flat ceiling; the weak case (fgas$+2\sigma$) shows no ceiling at either radius.}
\label{fig:entropy_2r200m}
\end{figure*}

\textit{Appendix: Entropy scaling beyond the virial radius.}---Figure~\ref{fig:entropy_2r200m} extends the entropy measurements of Fig.~\ref{fig:entropy} to $2\,r_\mathrm{200m}$. In the jet case, over the mass range $10^{12.5}$--$10^{13.5}\,\Msol$ the entropy slope $K \propto M^{0.22}$ is unchanged between $r_\mathrm{200m}$ and $2\,r_\mathrm{200m}$, confirming that the angular structure of the heated region is preserved during outward transport. The fiducial thermal case, by contrast, develops a positive slope at $2\,r_\mathrm{200m}$ that is absent at $r_\mathrm{200m}$. At low masses, the jet outflow entropy now exceeds the thermal, reversing sign relative to $r_\mathrm{200m}$. We expect that only the highest entropy gas in the system penetrates to $2\,r_\mathrm{200m}$. In the jet case, this is the directly processed material within the broadened cone. In the thermal case, the shock heats gas more isotropically to a lower peak entropy per particle. This gas stalls between $r_\mathrm{200m}$ and $2\,r_\mathrm{200m}$, and the flat ceiling gives way to a mass-dependent slope as lower-mass halos lose their outflows first. The thermal feedback variants confirm this interpretation: the strong AGN case (fgas$-8\sigma$) maintains a nearly flat ceiling even at $2\,r_\mathrm{200m}$, consistent with higher-entropy gas remaining buoyant over a broader radial range, while the weak case (fgas$+2\sigma$) shows no ceiling at either radius.
The thermal and jet cases therefore converge in scaling behavior beyond the virial radius, with the difference at $r_\mathrm{200m}$ arising from the geometry of the initial energy deposition rather than the energetics.

\end{document}